\documentclass{aa}  
\usepackage{natbib}
\bibpunct{(}{)}{;}{a}{}{,} 

\usepackage{soul}
\usepackage{graphicx}
\usepackage{txfonts}
\usepackage{url}
\usepackage{color}
\usepackage{natbib}
\usepackage{rotating}
\usepackage{lineno}
\usepackage{xspace}
\usepackage{paralist}

\newcommand{\pmn}{PMN\,J1603$-$4904\xspace}
\newcommand{\xmm}{\textsl{XMM-Newton}\xspace}
\newcommand{\suz}{\textsl{Suzaku}\xspace}

\newcommand{\fe}{neutral Fe~K$\alpha$ line\xspace}

\begin{document}

\title{A redshifted Fe\,K$\alpha$ line from the unusual $\gamma$-ray
  source \pmn}

\author{C. M\"uller\inst{\ref{affil:wuerzburg},\ref{affil:remeis}}
 \and F.~Krau\ss\inst{\ref{affil:wuerzburg},\ref{affil:remeis}}
   \and T.~Dauser\inst{\ref{affil:remeis}}
 \and A.~Kreikenbohm\inst{\ref{affil:wuerzburg},\ref{affil:remeis}}
    \and T.~Beuchert\inst{\ref{affil:wuerzburg},\ref{affil:remeis}}
 \and M.~Kadler\inst{\ref{affil:wuerzburg}}
\and R.~Ojha\inst{\ref{affil:umbc},\ref{affil:nasa_gsfc}, \ref{affil:cua}}
 \and J.~Wilms\inst{\ref{affil:remeis}}
 \and M.~B\"ock\inst{\ref{affil:wuerzburg},\ref{affil:remeis},\ref{affil:mpifr}}
\and B.~Carpenter\inst{\ref{affil:nasa_gsfc},\ref{affil:cua}}
 \and M.~Dutka\inst{\ref{affil:nasa_gsfc}}
\and A.~Markowitz\inst{\ref{affil:ucsd},\ref{affil:remeis} }
\and W.~McConville\inst{\ref{affil:nasa_gsfc}}
\and K.~Pottschmidt\inst{\ref{affil:umbc},\ref{affil:nasa_gsfc}}
 \and \L{}.~Stawarz\inst{\ref{affil:jaxa},\ref{affil:krakow}}
 \and G.~B.~Taylor\inst{\ref{affil:unm}}
            }

   \institute{
    Institut f\"ur Theoretische Physik und Astrophysik, Universit\"at
  W\"urzburg, Am Hubland, 97074 W\"urzburg, Germany\\
  \email{cornelia.mueller@astro.uni-wuerzburg.de} \label{affil:wuerzburg}
  \and
  Dr. Remeis Sternwarte \& ECAP, Universit\"at Erlangen-N\"urnberg,
  Sternwartstrasse 7, 96049 Bamberg, Germany \label{affil:remeis}
  \and
  Center for Space Science and Technology, University of Maryland
  Baltimore County, Baltimore, MD 21250, USA \label{affil:umbc}
 \and
  CRESST and NASA, Goddard Space Flight Center, Astrophysics Science
  Division, Code 661, Greenbelt, MD 20771, USA  \label{affil:nasa_gsfc}
 \and 
  Catholic University of America, Washington, DC 20064, USA \label{affil:cua}
  \and
  Max-Planck-Institut f\"ur Radioastronomie, Auf dem H\"ugel 69, 53121
  Bonn, Germany 
  \label{affil:mpifr}
  \and
  University of California, San Diego, Center for Astrophysics and
  Space Sciences, 9500 Gilman Drive, La Jolla, CA 92093-0424,
  USA \label{affil:ucsd}
  \and Institute of Space and Astronautical Science JAXA, 3-1-1
  Yoshinodai, Chuo-ku, Sagamihara, Kanagawa 252-5210,
  Japan \label{affil:jaxa}
  \and 
  Astronomical Observatory, Jagiellonian University, ul. Orla 171,
  Krak\'ow 30-244, Poland \label{affil:krakow}
 \and
  Department of Physics and Astronomy, University of New Mexico,
  Albuquerque, NM 87131, USA \label{affil:unm}
             }

   \date{\today}

  \abstract
   {Multiwavelength observations have revealed the highly unusual properties of the
$\gamma$-ray source \pmn, which are difficult to reconcile with any
other well established
$\gamma$-ray source class. The object is either a very atypical blazar
or compact jet source seen at a larger angle to the line of sight.}
   {In order to determine the physical origin of the high-energy
     emission processes in \pmn, we study the X-ray spectrum in detail.}
   {We performed quasi-simultaneous X-ray observations with \xmm and
     \suz in 2013 September, resulting in the first high
     signal-to-noise X-ray spectrum of this source. }
   {The 2--10\,keV X-ray spectrum can be well described by an
     absorbed power law with an emission line at $5.44\pm0.05$\,keV
     (observed frame). Interpreting this feature as a K$\alpha$ line
     from neutral iron, we determine the redshift of \pmn
     to be \mbox{$z=0.18\pm0.01$}, corresponding to a luminosity distance
     of $872\pm54$\,Mpc.} 
  {The detection of a redshifted X-ray emission line further
    challenges the original BL\,Lac classification of \pmn. This
    result suggests that the source is observed at a larger angle to
    the line of sight than expected for blazars, and thus the source
    would add to the elusive class of $\gamma$-ray loud misaligned-jet
    objects, possibly a $\gamma$-ray bright young radio galaxy.}

   \keywords{Galaxies: active -- Galaxies: jets -- Galaxies: individual:
     PMN\,J1603$-$4904  -- X-rays: galaxies -- Gamma rays: galaxies}

   \maketitle
%
\section{Introduction}\label{sec:intro}
Active Galactic Nuclei (AGN) are among the brightest extragalactic
sources at X-ray energies. According to the unification model
\citep{Antonucci1993,Urry1995}, the orientation of AGN with respect to
the line of sight, different accretion rates, and the presence of a
powerful jet lead to different spectral appearances. Jet-dominated
sources at the smallest jet inclination angles are classified as
blazars. Sources which are oriented at larger angles to the line of
sight are typically less dominated by beamed jet emission. The origin
of the X-ray emission of radio-loud AGN is still an open question
\cite[e.g.,][]{Fukazawa2014}. It can arise from Comptonization of
seed photons from the accretion disc in the hot corona, synchrotron or
inverse Compton emission, or a combination of these. The most prominent
emission line in AGN X-ray spectra is the Fe~K$\alpha$ line at
$\sim$6.4\,keV \citep[e.g.,][]{Nandra2006,Shu2010,Ricci2014}. This
line is usually explained by fluorescence of neutral or mildly ionized
iron produced by the primary, hard X-ray photons irradiating the
accretion disk or the molecular torus. Measurements of iron line emission
probe therefore the inner regions of AGN. The detection of an iron
line is a clear indication of reprocessed radiation from matter in the
vicinity of the primary X-ray source. Iron lines are common features
in spectra of radio-quiet and radio-loud sources at larger jet angles
to the line of sight \citep{Bianchi2004}. In contrast, typical blazar
X-ray spectra are featureless continua, clearly
dominated by the beamed jet emission that outshines any putative
underlying emission components \citep[e.g.,][]{Rivers2013,Grandi2006}.

In \citet[][hereafter Mue14]{Mueller2014a}, we discussed the
multiwavelength properties of \pmn, which is associated with a bright
$\gamma$-ray source with a remarkably hard spectrum \citep[\object{2FGL\,J1603.8--4904};][]{2fgl}.
These properties are difficult to reconcile with its original
classification as a BL\,Lac type blazar \citep{2fgl,1fhl}: High
resolution TANAMI \citep{Ojha2010a} radio observations using Very Long Baseline
Interferometry (VLBI) revealed a symmetric structure on
milliarcsecond (mas) scales with the brightest, most compact component
at the center. The broadband spectral energy distribution (SED) showed
an excess in the infrared, which could be modeled with a blackbody
spectrum with $T\sim1600$\,K, consistent with the emission of
circumnuclear dust heated by the disk photons. Such thermally reprocessed radiation is
not expected in typical blazar SEDs. Optical measurements constrained the redshift
to $z\lesssim 4.24$ \citep{Shaw2013a}. Based on the results presented
by Mue14, however, no conclusive classification of \pmn could be made.
While \textsl{Swift}/XRT observations showed a faint X-ray
counterpart, low photon statistics did not allow us to sufficiently
constrain its X-ray spectrum.

In this letter, we report on the first \xmm and \suz observations of \pmn conducted
within the scope of the TANAMI multiwavelength program
(Sect.~\ref{sec:obs}). In Sect.~\ref{sec:results} we present the X-ray
spectral characteristics and discuss its implications in
Sect.~\ref{sec:discussion}. 

\section{Observations and data reduction}\label{sec:obs}

In 2013 September, we performed quasi-simultaneous X-ray observations
of \pmn with \textsl{XMM-Newton} \citep[][obs-id 0724700101, performed
  2013-09-17]{Jansen2001} and \textsl{Suzaku} \citep[][obs-id
  708035010, performed 2013-09-13]{Koyama2007} in order to constrain
the X-ray spectrum better than previous observations with
\textsl{Swift}/XRT \citep[][Mue14]{Burrows2005}. The
\textsl{XMM-Newton} data processing, analysis, and extraction of the
spectrum were performed using the standard tasks of the \textit{XMM
  System Analysis Software} (SAS 13.5.0). We detected a single,
unresolved, X-ray source at the radio position of \pmn. Only data from
EPIC-PN \citep[][exposure time 39.0\,ks]{strueder:01a} and
EPIC-MOS\,1 \citep[][exposure time 33.4\,ks]{turner:01a} cameras
were used, as the usable exposure of EPIC-MOS\,2 was only 4.7\,ks. The
PN spectrum was extracted using a circle with $35''$ radius centered
on the coordinates of the radio source. For the background spectrum a
circular source-free region of $50''$ radius was chosen. For the
MOS\,1 data extraction, background and source regions with radii of
$100''$ were used. 
We chose larger extraction radii for the background to minimize the measurement uncertainty.
We used the HEASOFT v6.15 package for the
\textsl{Suzaku} data analysis. Extraction and background regions of
$94''$ were used for all XIS detectors, which after filtering had an
exposure time of 48.5\,ks. The spectral analysis was performed using
the \textit{Interactive Spectral Interpretation System} \citep[ISIS,
  Version 1.6.2-27,][]{Houck2000}. An independent analysis of the \xmm
and \suz data revealed no significant flux variability during the
observed period. In order to improve the statistics and ensure the
validity of the $\chi^2$-statistics, all \mbox{1--10\,keV} spectra were
rebinned to a minimum signal-to-noise ratio of 5 and modeled
simultaneously considering the background. Throughout this paper we use the standard cosmological
model with
$\Omega_\mathrm{m}=0.3$, $\Lambda = 0.7$, $H_0 =
70\mathrm{\,km\,s^{-1}\,Mpc^{-1}}$ \citep{Beringer2012}.
All uncertainties quoted in the following are given at the 90\% confidence.

\section{Results}\label{sec:results}
\subsection{X-Ray continuum}\label{sec:analysis}
\begin{figure}
    \includegraphics[width=\hsize]{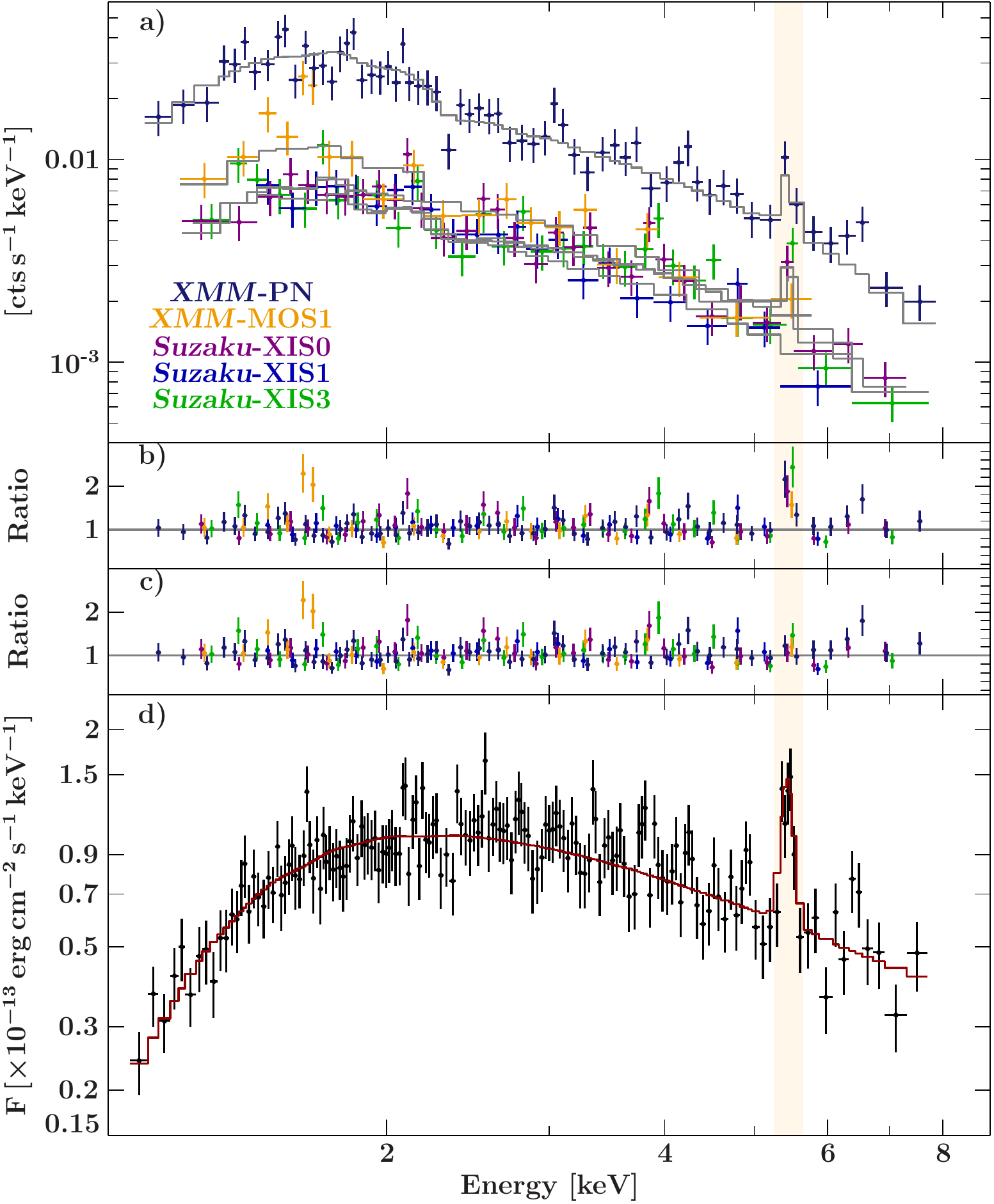}
    \caption{Simultaneous fit to the \textsl{XMM-Newton} and
      \textsl{Suzaku} data (see Table~\ref{table:pmnbestfit}). The
      data are best fitted by an absorbed power-law component with an
      emission line at $\sim5.44\pm0.05$\,keV. \textit{(a):} Count
      spectrum for all detectors fitted with an absorbed power law and
      a Gaussian emission line (models shown in gray). Ratio of data
      to model for a fit of an absorbed power law (\textit{b}) and
      including a Gaussian emission line (\textit{c}). \textit{(d):}
      Unfolded, combined spectrum of all data sets with the best-fit
      model (red). The shaded region highlights the position of the
      emission line.}
    \label{fig:pmnxray}
\end{figure}

The quasi-simultaneous data from \textsl{XMM}/PN, MOS\,1, and
\textsl{Suzaku}/XIS were simultaneously modeled with an absorbed
power-law component, a Gaussian emission line, and cross-calibration
constants to account for the relative flux calibrations of the
instruments. We use the $\mathtt{tbnew}$ model \citep{Wilms2012} to
account for neutral Galactic absorption \citep[Galactic
H\textsc{I} value is fixed to $N_\mathrm{H} = 6.32 \times
  10^{21}\,\mathrm{cm}^{-2}$;][]{Kalberla2005}, and the cross sections
and abundances of \citet{Verner1996} and \citet{Wilms2000},
respectively. 
Further residuals reveal the presence of source intrinsic absorption,
which we model with an additional, redshifted absorption component
($\mathtt{tbnew\_z}$, see below).
We detect a strong emission line that can be modeled by
a Gaussian component at $5.44\pm0.05$\,keV with an equivalent width of
$EW=200\pm90$\,eV (Fig.~\ref{fig:pmnxray}a).
Using the centroid energy of the Gaussian emission line with respect
to the \fe line rest frame energy, we can constrain the redshift of
the system to $z=0.18\pm0.01$ (see Sect.~\ref{sec:line} for details).
 We find a photon index of
$\Gamma_\mathrm{X}=2.07^{+0.04}_{-0.12}$ with an intrinsic absorption of
\mbox{$N_\mathrm{H} = 2.05^{+0.14}_{-0.12}\times
  10^{22}\,\mathrm{cm}^{-2}$}. The best fit model parameters can be
found in Table~\ref{table:pmnbestfit}. Adding a Gaussian line to
describe the emission line improves the fit from $\chi^2=198.1$ (165
degrees of freedom [dof]) to $\chi^2=183.3$ (162\,dof).
To test for the significance of the line feature at 5.44\,keV, we
performed a Monte Carlo simulation based on the Null-Hypothesis that the
intrinsic model for the measured counts is an absorbed power-law lacking
a line. This hypothesis is rejected with a p-value of $p<1.1\times10^{-4}$.
Figure~\ref{fig:pmnxray} shows the spectra of all single datasets with
the best fit model and the combined datasets regridded to a joint
energy grid.

\subsection{Iron emission line}\label{sec:line}                
Due to its high fluorescent yield, the Fe K$\alpha$ transition
produces the most prominent line feature in the 2--10\,keV energy
range. Such lines are seen in the X-ray spectra of the majority of
Seyfert and radio galaxies
\citep{delacalle2010}. In most sources where redshifts are known \citep[e.g.,][]{Markowitz2007,Shu2010,Ricci2014}, the
line energy and equivalent width are consistent with an origin by
fluorescence with rest-frame energy of $\sim$6.4\,keV in a neutral to mildly ionized medium that is irradiated
by hard X-rays.
Even in sources
showing relativistically broadened Fe K$\alpha$ lines from the
innermost regions of the accretion disk, narrow line components
consistent with a rest-frame energy of 6.4\,keV are typically seen 
\citep{Risaliti2013,Marinucci2014,Marinucci2014_mcg}.
We therefore
consider it most probable that the emission line seen in \pmn at
$5.44\pm0.05$\,keV is also emitted intrinsically at 6.4\,keV\footnote{In the very unlikely event that the line is instead due to emission from a
strongly ionized plasma, for hydrogen-like iron, the Ly$\alpha$
resonance at 6.966\,keV would yield a redshift of $z = 0.28\pm0.01$.}, such
that the corresponding redshift of the source can be constrained to $z =
0.18\pm0.01$.

Adopting this redshift, the luminosity distance is $872\pm54$\,Mpc.
In the rest-frame, the intrinsic X-ray luminosity, neglecting Galactic
and source intrinsic absorption, is 
$L_\mathrm{2\text{--}10\,keV}=(4.03\pm0.20)\times 10^{43}\,\mathrm{erg\,s^{-1}}$ 
and the $\gamma$-ray luminosity is
$L_\mathrm{1\text{--}100\,GeV}=(8.1\pm1.1)\times 10^{45}\mathrm{\,erg\,s^{-1}}$ \citep[for
  \mbox{$S_\mathrm{1\text{--}100\,GeV}=(1.44\pm0.07)\times
    10^{-10}\,\mathrm{erg\,s^{-1}\,cm^{-2}}$},][]{2fgl}.
At 8.4\,GHz, we derive a rest-frame luminosity density of
$L_\mathrm{8.4\,GHz}=(7.3\pm1.5)\times 10^{32}\,\mathrm{erg 
  s^{-1}\,Hz^{-1}}$ 
(based on the flux from Mue14). The comparison of ATCA
observations at arcsec resolution and VLBI observations with
milliarcsecond resolution shows that $\sim$80\% of the radio emission
is emitted on scales of $\sim$15\,mas (Mue14), i.e., $\sim$46\,pc at
$z=0.18$. The remaining $\sim$20\% is emitted on larger scales which
are constrained to $\lesssim 1''$ (Mue14), i.e., $\sim$3\,kpc at $z=0.18$.

\begin{table}
 \caption{Best fit parameters using an absorbed power-law model with a
   Gaussian emission line.}\label{table:pmnbestfit}      
    \centering    
\begin{tabular}{ll} 
 \hline\hline
Parameter  &  Value \\
\hline
$N_\mathrm{H}$ (intrinsic) &  $2.05^{+0.14}_{-0.12}\times10^{22}\,\mathrm{cm}^{-2}$  \\
$F_\mathrm{2\text{--}10\,keV}$ (de-absorbed) &
$\left(4.39\pm0.17\right)\times
10^{-13}\,\mathrm{\,erg\,s^{-1}\,cm^{-2}}$ \\ 
$\Gamma$   &   $2.07^{+0.04}_{-0.12}$    \\
$E_\mathrm{Fe\ K\alpha}$    &    $5.44\pm0.05$\,keV    \\ 
$\sigma_\mathrm{Fe\ K\alpha}$  &$<0.06$\,eV    \\ 
$EW$ &  $200\pm90$eV \\
$\mathrm{const_{pn}}$ &   1.0   \\ 
$\mathrm{const_{MOS1}}$   &  $1.00^{+0.09}_{-0.08}$     \\ 
$\mathrm{const_{XIS0}}$   &  $1.57\pm0.11$    \\ 
$\mathrm{const_{XIS1}}$   &   $1.46\pm0.12$   \\ 
$\mathrm{const_{XIS3}}$   &    $1.63\pm0.12$  \\ 
\hline
$\chi^{2}/\mathrm{dof}$ &  183.0/162   \\  
\hline\hline
\end{tabular}
\end{table}


\section{Discussion}\label{sec:discussion}
The X-ray properties derived from the \xmm and \suz spectra allow us
to further constrain the classification of this unusual $\gamma$-ray
source. Here we compare the X-ray spectrum of \pmn to the typical
spectra of different radio-loud AGN (see Table~\ref{table:compare}).

Compared to other source types, blazars are the most luminous AGN at
the X-ray energies \citep[e.g.,][]{Chang2010PhDT,Rivers2013}. With an X-ray
luminosity of $L_\mathrm{2\text{--}10\,keV}\sim
4\times10^{43}\,\mathrm{erg\,s^{-1}}$, \pmn is only comparable to
the fainter blazars of the BL\,Lac type. Its $\gamma$-ray luminosity
of $L_\gamma\sim8\times 10^{45}\mathrm{\,erg\,s^{-1}}$ is comparable
to low or intermediate peaked\footnote{
The synchrotron peak frequency
derived from the parametrization of the broadband spectral energy
distribution is $\nu_\mathrm{sync} \simeq 2.2\times10^{12}$\,Hz
(Mue14), which would be more in line with a low-peaked object \citep[defined by $\nu^S_\mathrm{peak}<10^{14}$\,Hz;][]{Fossati1998,Donato2001}.} 
BL\,Lac objects and flat
spectrum radio quasars \citep{2lac}. Since
blazar X-ray spectra are dominated by beamed jet emission, they are
generally featureless\footnote{As a rare exception, \citet{Grandi2004}
  discuss the detection of a narrow Fe\,K$\alpha$ line in the
  flat-spectrum radio quasar \object{3C\,273}. In this case the thermal
  (unbeamed) and the beamed jet emission could be disentangled due to
  source variability, suggesting the presence of an underlying
  Seyfert-like component \citep{Soldi2008}.} and can in most cases be modeled with a
simple or broken power law \citep{Chang2010PhDT,Ushio2010,Rivers2013}. The
strong iron line in the X-ray spectrum of \pmn therefore suggests a
non-blazar nature. The intrinsic X-ray absorption
($N_\mathrm{H}\sim2\times 10^{22}\,\mathrm{cm^2}$) is furthermore
unusually high compared to other blazars
\citep{Kubo1998,Chang2010PhDT,Rivers2013}. We therefore conclude that
its original classification as a BL\,Lac \citep{2fgl,Shaw2013a} is
very unlikely.

The multiwavelength properties of \pmn opened the door to an
alternative classification as a jet source seen at a larger
inclination angle, i.e., a radio galaxy (Mue14). Overall, the X-ray
properties of \pmn are comparable to those of radio galaxies
(Table~\ref{table:compare}). In about 50\% of all broad-line radio
galaxies, Fe~K$\alpha$ lines with equivalent widths of typically a few
100\,eV are detected \citep[e.g.,][]{Sambruna1999}.  \pmn has a
comparable equivalent width of $EW=200\pm90$\,eV.  We caution,
however, that only few non-blazar sources are detected at $\gamma$-ray
energies \citep{Abdo2010_misaligned,2fgl,Katsuta2013}.  The
$\gamma$-ray luminosities of such ``misaligned'' jets show a broad
range from $L_\gamma\sim 10^{41}$--$10^{46}\mathrm{\,erg\,s^{-1}}$,
with FR~I galaxies being typically less $\gamma$-ray luminous than
FR~II sources. The $\gamma$-ray luminosity of \pmn is only comparable
to the most powerful misaligned sources \citep{Abdo2010_misaligned}.

Among the sources seen at a large inclination angle, Compact Steep
Spectrum (CSS) and Gigahertz Peaked Spectrum (GPS) sources can be
interpreted as the younger versions of evolved radio galaxies, often
showing compact symmetric object (CSO) radio morphologies at parsec
scales \citep{Readhead1996b,Readhead1996a,Odea1998}. Owing to its
milliarcsecond-scale properties, we discussed the possible
classification of \pmn as a CSO in Mue14.  The 2--10\,keV spectra of
GPS and CSO sources can generally be well modeled by absorbed
power laws
\citep[e.g.,][]{Vink2006,Guainazzi2006,Siemiginowska2008,Tengstrand2009,KunertBajraszewska2014}.
Fe K$\alpha$ emission is only detected in a few sources
\citep[e.g.,][]{Guainazzi2006,Siemiginowska2009,Tengstrand2009}. The
X-ray properties of \pmn are a good match to the findings in these
sample studies of GPS and CSS sources. These sources generally have a
radio luminosity density of
$L\gtrsim10^{32}$--$10^{36}\,\mathrm{erg\,s^{-1}\,\,Hz^{-1}}$ 
at 5\,GHz \citep[][for combined complete
  samples]{OdeaBaum1997}. Thus \pmn is compatible with the less
powerful GPS and CSS sources. Its linear size is comparable to the
canonical size limit for CSOs ($<1$\,kpc) or medium-size symmetric
objects \citep[$<15$\,kpc;][]{Fanti1995,Readhead1995,Readhead1996a}.
Gamma-ray emission was predicted from theoretical modeling of GPS
sources
\citep[e.g.,][]{Stawarz2008,Ostorero2010,Kino2013,Migliori2014} and
first source candidates have been discussed, e.g., \object{4C\,+55.17}
\citep{McConville2011} with similar broadband properties as \pmn. If
confirmed as a $\gamma$-ray loud young radio galaxy, then \pmn would
be a well-suited object to investigate the origin of high-energy emission.


\begin{table*}
\caption{Typical high-energy spectral parameters for different
  source classes compared to results for \pmn.} 
    \label{table:compare}      
    \centering
\begin{tabular}{c|ccc|c}
\hline
  & Blazars & Radio Galaxies & GPS/CSS & \pmn \\
\hline\hline
$L_\mathrm{2\text{--}10keV}$ [$\mathrm{erg\,s^{-1}}$] &
$\sim$$10^{43}$--$10^{46}$\tablefootmark{(a)}  &
$\sim$$10^{42}$--$10^{45}$\tablefootmark{(b)} &  
$10^{42}$--$10^{46}$\tablefootmark{(c)} &$(4.03\pm0.20)\times10^{43}$ \\ 
$L_\mathrm{\gamma}$ [$\mathrm{erg\,s^{-1}}$] &
  $10^{44}$--$10^{49}$\tablefootmark{(d)} 
  &$\sim10^{41}$--$10^{46}$\tablefootmark{(d)} &--
  &$(8.1\pm1.1)\times10^{45}$  \\ 
$\Gamma_\mathrm{2\text{--}10\,keV}$ &$1.4$--$2.5$\tablefootmark{(e)} &
1.7--1.8\tablefootmark{(f)}
 & 1.7--2.0\tablefootmark{(g)} & $2.07_{-0.12}^{+0.04}$\\ 
$N_\mathrm{H}$ (intrinsic) [$\mathrm{cm^{-2}}$] &
$\lesssim$$10^{21}$\tablefootmark{(e)} & 
$\sim$$10^{21}$--$10^{24}$\tablefootmark{(f)} &
$\sim$$10^{22}$\tablefootmark{(h)} &
$2.05^{+0.14}_{-0.12}\times10^{22}$\\
Fe K$\alpha$ line & no\tablefootmark{(e)}  & yes\tablefootmark{(e)}
& yes\tablefootmark{(i)}  & yes \\ 
$EW_\mathrm{Fe\,K\alpha}$ [eV] & -- & ~\tablefootmark{(e)} & $\lesssim 150$\tablefootmark{(i)} & $200\pm90$\\
\hline
\end{tabular}
\tablefoot{ 
\tablefoottext{a}{\citet[][for the radio flux density-limited MOJAVE~1 sample]{Chang2010PhDT}}
\tablefoottext{b}{\citet{Sambruna1999}}
\tablefoottext{c}{\citet{Vink2006}, \citet{Siemiginowska2008}.}
\tablefoottext{d}{\citet{Abdo2010_misaligned}, \citet{2lac}.}
\tablefoottext{e}{\citet{Sambruna1999}, \citet{Chang2010PhDT}, \citet{Rivers2013}.}
\tablefoottext{f}{
  \citet{Sambruna1999}, \citet{Eracleous2000}, \citet{Evans2006},
  \citet{Hardcastle2006}, \citet{Grandi2006}, \citet{Rivers2013}.} 
\tablefoottext{g}{\citet{Siemiginowska2008}.}
\tablefoottext{h}{\citet{Vink2006}, \citet{Guainazzi2006},
  \citet{Siemiginowska2008}, \citet{Tengstrand2009}, \citet{KunertBajraszewska2014}.}
\tablefoottext{i}{\citet{Guainazzi2006}, \citet{Siemiginowska2009}, \citet{Tengstrand2009}.}
}
\end{table*}

\section{Conclusions}\label{sec:conclusion}
We have presented new X-ray observations of \pmn, which further
challenge its previous classification as a blazar and strongly suggest
that this jet system is seen at a larger inclination angle to the line
of sight. Its X-ray spectrum can be modeled by an absorbed power law
and a Gaussian emission line at $5.44\pm0.05$\,keV, and can be best
explained as emission from a non-blazar source. These X-ray properties
are consistent with a young or evolved radio galaxy. We interpret the
significantly detected spectral line as a \fe with rest frame energy
6.4\,keV. This identification results in the first redshift
measurement for this source of $z=0.18\pm0.01$, i.e., a luminosity
distance of $872\pm54$\,Mpc.

\pmn is associated with a hard spectrum $\gamma$-ray source
\citep{2fgl}\footnote{Note that a small possibility of a false
  association of the $\gamma$-ray source with \pmn still remains,
  although in such a case an even more exotic explanation for the
  $\gamma$-ray origin would be
  required.}. If it is indeed seen at a larger inclination angle where
the observed emission is less intensified by Doppler boosting than in
blazars, it adds to the class of so-called misaligned sources rarely
detected at $\gamma$-ray energies \citep{Abdo2010_misaligned}, with a
remarkably high $\gamma$-ray luminosity.

Further multiwavelength observations are required to confirm \pmn as a
$\gamma$-ray loud young radio galaxy: VLBI monitoring will allow us to
check for proper motion in opposite directions with respect to the
core. Low radio frequency observations could help to determine a
putative peak at $\lesssim 1\,$GHz, thus increasing confidence in its identification as
a GPS source, and to constrain the extended radio emission.


\begin{acknowledgements}
We thank the referee for the helpful comments, and 
R.~Schulz and P.G.~Edwards for the useful discussions that improved
the manuscript.
C.M.~acknowledges the support of the Bundesministerium f\"ur
Wirtschaft und Technologie (BMWi) through Deutsches Zentrum f\"ur
Luft- und Raumfahrt (DLR) grant 50\,OR\,1404 and of the
Studienstiftung des Deutschen Volkes. This research was funded in part
by the National Aeronautics and Space Administration (NASA) through
Fermi Guest Investigator grants NNH09ZDA001N, NNH10ZDA001N, and
NNH12ZDA001N, by BMWi through DLR grant 50\,OR\,1311, by Deutsche
Forschungsgemeinschaft grant WI1860/10-1, and by an appointment to the
NASA Postdoctoral Program at the Goddard Space Flight Center,
administered by Oak Ridge Associated Universities through a contract
with NASA. We thank J.~E.~Davis for the development of the
\texttt{slxfig} module that has been used to prepare the figure in
this work. This research has made use of ISIS functions provided by
ECAP/Dr.\,Karl Remeis-Observatory (Bamberg, Germany)
and MIT (http://www.sternwarte.uni-erlangen.de/isis/).
\end{acknowledgements}

\bibliographystyle{aa}
\bibliography{aaabbrv,pmn}

\end{document}